\documentclass[iop]{emulateapj}

\usepackage{amsmath}    % Fancy math fonts
\usepackage{graphicx}   % Include figure files
\usepackage{dcolumn}    % Align table columns on decimal point
\usepackage{bm}         % Bold math
\usepackage{lineno}     % Line numbering package
\usepackage{hyperref}   % Hyperlinks in references
\usepackage{booktabs}  
\usepackage{xspace}     % Properly inserts spaces for math-mode definitions
\usepackage[usenames,dvipsnames]{color}
\shorttitle{Cosmic Ray Anisotropy with IceTop}
\shortauthors{IceCube Collaboration}

\def \mus      {$\mu\text{s}$\xspace}                       % Micro-second                                                                                                        
\def\eq#1{\begin{linenomath}\begin{equation}#1\end{equation}\end{linenomath}}

\begin{document}
  
% Frontmatter goes here: title, author list, abstract, and keywords
% -----------------------------------------------------------------------------
\title{Observation of Cosmic Ray Anisotropy with the IceTop Air Shower Array}

\author{
IceCube Collaboration:
M.~G.~Aartsen\altaffilmark{1},
R.~Abbasi\altaffilmark{2},
Y.~Abdou\altaffilmark{3},
M.~Ackermann\altaffilmark{4},
J.~Adams\altaffilmark{5},
J.~A.~Aguilar\altaffilmark{6},
M.~Ahlers\altaffilmark{2},
D.~Altmann\altaffilmark{7},
K.~Andeen\altaffilmark{2},
J.~Auffenberg\altaffilmark{2}
% ...
}
\altaffiltext{1}{School of Chemistry \& Physics, University of Adelaide, Adelaide SA, 5005 Australia}
\altaffiltext{2}{Dept.~of Physics and Wisconsin IceCube Particle Astrophysics Center, University of Wisconsin, Madison, WI 53706, USA}
\altaffiltext{3}{Dept.~of Physics and Astronomy, University of Gent, B-9000 Gent, Belgium}
\altaffiltext{4}{DESY, D-15735 Zeuthen, Germany}
\altaffiltext{5}{Dept.~of Physics and Astronomy, University of Canterbury, Private Bag 4800, Christchurch, New Zealand}
\altaffiltext{6}{D\'epartement de physique nucl\'eaire et corpusculaire, Universit\'e de Gen\`eve, CH-1211 Gen\`eve, Switzerland}
\altaffiltext{7}{Institut f\"ur Physik, Humboldt-Universit\"at zu Berlin, D-12489 Berlin, Germany}
% ...

\begin{abstract}
We report on the observation of anisotropy in the arrival direction
distribution of cosmic rays at PeV energies.  The analysis is based
on data taken between 2009 and 2012 with the IceTop air shower array 
at the South Pole. IceTop, an integral part of the IceCube detector,
is sensitive to cosmic rays between 100 TeV and 1 EeV.  With the current
size of the IceTop data set, searches for anisotropy at the $10^{-3}$
level can, for the first time, be extended to PeV energies.  We divide 
the data set into two parts with median energies of 400\,TeV and 2\,PeV, 
respectively.  In the low energy band, we observe a strong deficit with 
an angular size of about $30^{\circ}$ and an amplitude of 
$(-1.58 \pm 0.46_{\mathrm{stat}} \pm 0.52_{\mathrm{sys}}) \times 10^{-3}$ 
at a location consistent with previous observations of cosmic rays with 
the IceCube neutrino detector.  The study of the high energy band shows 
that the anisotropy persists to PeV energies and increases in amplitude to 
$(-3.11 \pm 0.38_{\mathrm{stat}} \pm 0.96_{\mathrm{sys}}) \times 10^{-3}$.
\end{abstract}


% Article sections go here: intro, conclusions, bibliography, and parts between
% -----------------------------------------------------------------------------

%************************************************************************************
\section{Introduction and Motivation}\label{sec:Introduction}
%************************************************************************************

Over the last half-decade, several experiments in the northern and
southern hemisphere have reported anisotropy in the arrival direction
distribution of cosmic rays at TeV energies.  In the northern sky,
two features dominate the TeV cosmic ray sky: a large-scale structure with an amplitude of about 
$10^{-3}$ usually described as a dipole \citep{Munakata:1997ei,Amenomori:2005dy,Amenomori:2006bx,Guillian:2007, Abdo:2008aw}, and a small-scale structure with a 
few hot spots of angular size $10^{\circ}$ to 
$30^{\circ}$ \citep{Abdo:2008kr,Vernetto:2009xm}.

In the southern hemisphere, cosmic ray arrival directions at TeV energies
observed with the IceCube neutrino detector at the South Pole exhibit
features similar to those discovered in the northern sky.  In a data set with 
a median energy of 20\,TeV, the arrival direction distribution exhibits a
large-scale structure similar in orientation and shape to the large-scale feature
 observed in the northern sky \citep{Abbasi:2010mf}.  

In addition, there is a small-scale structure which is about a factor of five 
weaker in relative intensity than the large-scale structure.  This small-scale 
structure contains several regions of significant cosmic-ray excess and
deficit \citep{Abbasi:2011ai}.  

There are several models that can at least qualitatively explain the anisotropy.
Cosmic rays in this energy range are assumed to be accelerated in Galactic
sources, most likely in shocks from supernova explosions.  The transport of 
cosmic rays at these energies in the Galactic magnetic field is diffusive, 
and the flux from a single nearby source would be observed on Earth as a dipole 
with its maximum possibly oriented towards the source.  If a few supernova remnants from 
recent (10...100 kyrs) nearby supernovae were primarily responsible for the 
Galactic cosmic ray flux ~\citep{Erlykin:2006ri}, their combined flux on Earth
would be a superposition of these individual dipoles.  The observed large-scale
structure in the cosmic ray flux could be the sum of the contributions from 
a few nearby sources and from the large scale distribution of supernova 
remnants in our Galaxy \citep{Blasi:2011fm,Pohl:2012cc}.  Our limited knowledge of nearby 
supernova remnants renders a more quantitative explanation of the amplitude 
and the phase of the observed large-scale anisotropy impossible, but among
the predictions of this model is an increase of the anisotropy with
the energy of the primary cosmic rays.

The small-scale structure is more difficult to explain.  The Larmor
radius of a proton with 10\,TeV energy in a magnetic field with $\mu$G strength
(an estimate of the strength of magnetic fields in our Galaxy \citep{Bfield})
is only of order 0.01\,pc, so the observed small-scale anisotropy cannot 
correlate with nearby sources.  Recent studies link the smaller structure 
to cosmic ray progation in turbulent magnetic fields within a few tens of 
parsecs from Earth \citep{Giacinti:2011mz}.  In this case, the small-scale 
structure is also expected to show a dependence on energy.

Several experiments have studied the energy dependence of the anisotropy.  In the 
northern hemisphere, the EAS-TOP experiment reports anisotropy up to at least 400\,TeV.  
The data contain weak evidence for an increase in the amplitude of the anisotropy as 
a function of energy, as well as a change of phase \citep{Aglietta:2009mu}.
Measurements in the southern hemisphere with IceCube also indicate that 
anisotropy is still present in a data set with 400\,TeV median energy.  It 
differs in shape and strength from the anisotropy observed at 20\,TeV
and is no longer a superposition of large- and small-scale 
structures, but rather dominated by a single deficit region with an angular 
size of about $30^{\circ}$ \citep{Abbasi:2011zka}.

IceCube is primarily a neutrino detector designed to search 
for sources of astrophysical neutrinos.  The data set used in the cosmic 
ray analysis consists of downgoing atmospheric muons from cosmic ray air 
showers in the atmosphere above the IceCube detector.  The downgoing muons 
preserve the direction of the cosmic ray primary, but the muon energy is 
only a poor indicator of the air shower energy.  In addition, because of 
the high muon rate ($10^6$ times the neutrino rate in IceCube), these events 
are stored in a separate data format which only contains the results of a 
fast online reconstruction performed at the South Pole.  No raw data are 
preserved.  An offline reconstruction with more sophisticated algorithms 
to determine the muon energy is therefore not possible.  

Above 1\,TeV, the energy resolution is of order 0.5 in $\log(E)$ \citep{Abbasi:2011zka}.
The energy resolution is estimated as the standard deviation of the distribution of the difference 
$\log(E_{\mathrm{true}}) - \log(E_{\mathrm{reco}})$,
where $E_{\mathrm{true}}$ and $E_{\mathrm{reco}}$ are the true and 
reconstructed shower energies obtained from simulation studies, respectively.
This distribution has substantial tails, making it difficult to isolate a set of events with large median energy that is not 
contaminated by low-energy events. 

The IceTop air shower array is located at 2835\,m altitude on the surface 
of the ice sheet above the IceCube neutrino detector.  IceTop is a dedicated 
cosmic ray detector optimized for air shower observations at PeV energies.
IceTop is used to record not only the muonic component of the air showers,
but also the electromagnetic component, at ground level.  With the sparse sampling 
of the shower front typical for air shower arrays, it also has a considerably 
higher detection threshold for cosmic rays than IceCube.  The size and geometry 
of the array result in a threshold for reconstruction of air showers of 
approximately 300\,TeV.

As an air shower array, IceTop provides a more measured information per shower
than IceCube.  A study
of cosmic ray anisotropy with IceTop can therefore complement measurements
with IceCube.  With its high energy threshold, an energy resolution better than 0.1 in
$\log(E)$ \citep{Abbasi:2012nn}, and sensitivity to the cosmic ray composition, 
IceTop data are particularly useful for studying anisotropy at energies above 
1\,PeV.  Due to the lower data rate, the full event information is stored, and
a more careful offline reconstruction of the primary cosmic ray properties 
is possible.

The accumulated IceTop data set is not yet large enough for a detailed study 
of per-mille anisotropy in several energy bins.  However, the statistics collected 
between 2009 and 2012 are now sufficient to search for anisotropy in two energy 
bands centered at 400\,TeV and 2\,PeV.  The 400\,TeV data set can be compared to 
results based on downgoing muons in IceCube at a similar energy but for a
different cosmic ray composition model \citep{Abbasi:2011zka}.  
With the 2\,PeV data set, the search for anisotropy is extended to energies not 
previously explored.  In this paper, we report on the observation of anisotropy 
with IceTop at both energies.  

The paper is organized as follows.  In Section 2, we describe the 
IceTop detector and the data sample used in this analysis.  The 
analysis techniques and the results are briefly described in Section 3.  
The analysis is based on methods used in previous work; a more 
detailed description of these techniques can be found in \citet{Abbasi:2011ai}.  
Section 4 discusses systematic uncertainties, and Section 5 summarizes the paper.

%************************************************************************************
\section{Detector, Data Sets and Simulation}\label{sec:detector}
%************************************************************************************

%-------------------------------------
\subsection{The IceTop Detector}
%-------------------------------------

\begin{figure}[t]
  \begin{center}	
    \includegraphics[width=0.45\textwidth]{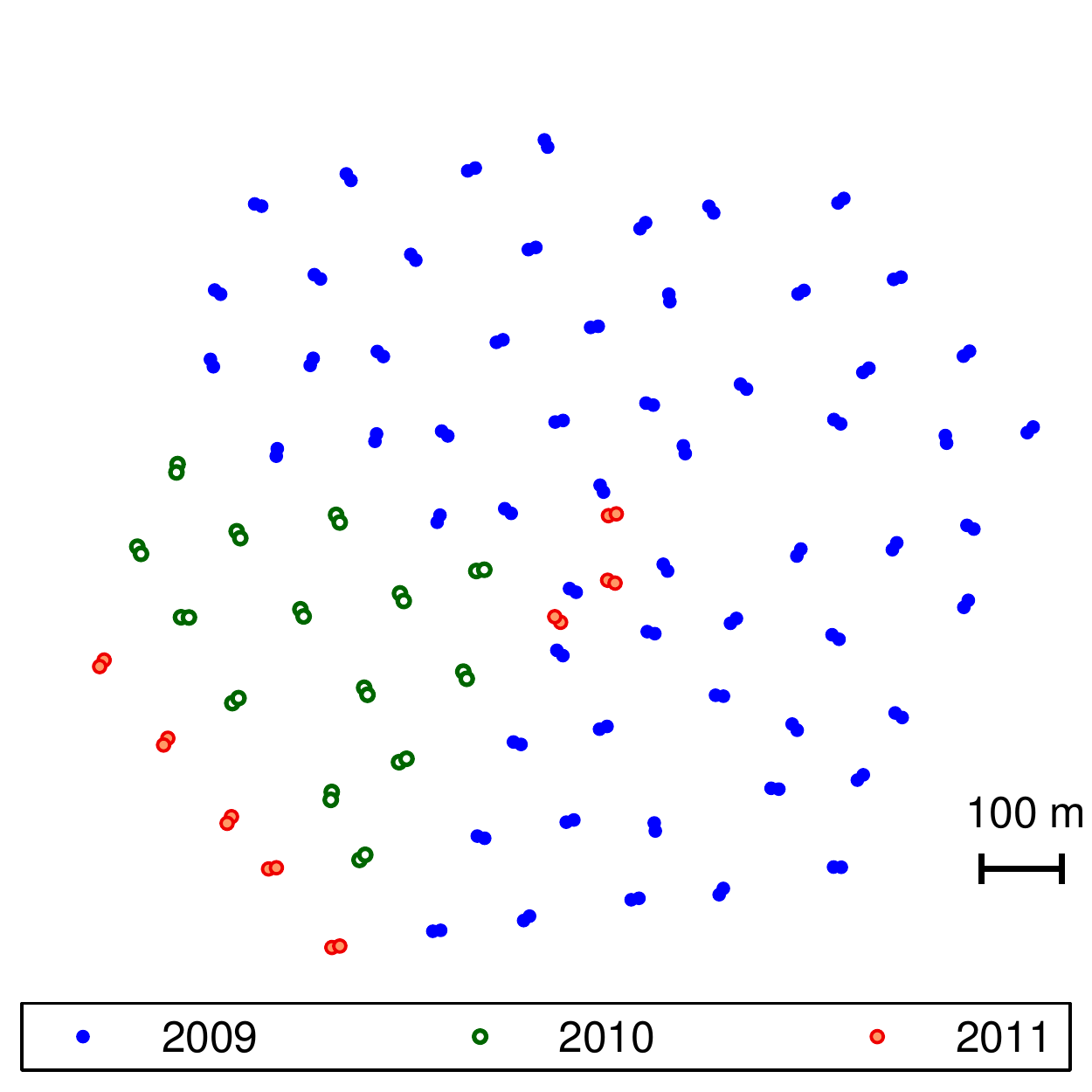}
    \caption{Detector configurations of the IceTop array, 2009-2011.  IT59 comprised
             59 stations deployed through January 2009 (blue circles).
             In 2009 and 2010 fourteen additional stations were deployed and
             the detector was operated in the IT73 configuration (blue
             and green circles).  The remaining eight stations (red circles) were deployed
             in late 2010.  The final IceTop configuration (IT81) consists of 81 stations and
             operated in 2011.}
    \label{fig:it_layout}
  \end{center}
\end{figure}

The IceTop cosmic ray air shower array consists of 81 stations distributed 
over an area of $1\,\mathrm{km}^2$ in a hexagonal grid with a distance 
of about 125\,m between neighboring stations.  During the construction phase 
of IceTop between 2005 and 2010, the detector was operated in several partial 
configurations.  In this work we use data taken during three periods: between
May 2009 and May 2010 when the detector was operated in a 59-station 
configuration (IT59); between May 2010 and May 2011 when IceTop operated with 73 
stations (IT73); and between May 2011 and May 2012 when the detector operated
in its final 81-station configuration (IT81).  The layout of these 
configurations is shown in Fig.~\ref{fig:it_layout}.  

IceTop is described in detail in \citet{Abbasi:2012nn}.  Each IceTop station is 
instrumented with two light-tight ice Cherenkov tanks separated by about 10\,m.  
Each tank is 1.8\,m in diameter, 1.3\,m in height, and is filled with transparent 
ice up to a height of 0.9\,m.  Frozen into the ice are two Digital Optical Modules 
(DOMs) that are used to detect the Cherenkov radiation emitted by charged leptons 
present in the cosmic ray air shower.  A DOM consists of a glass sphere that houses 
a 10" Hamamatsu photomultiplier tube (PMT) \citep{Abbasi:2010vc}, together with 
electronic boards used for filtering, digitization, and readout \citep{Abbasi:2008ym}.
 
The two DOMs inside each IceTop tank are operated at different PMT gains in
order to increase the dynamic range of the detector.  The high-gain DOMs in the
two tanks that form a station are run in local coincidence mode, and data
readout is enabled if one of the DOMs records photon hits within $\pm 1$ \mus 
of the other.  The trigger condition in IceTop requires at least six DOMs to have
recorded locally-coincident hits within a time window of 5 \mus, which
implies that at least two stations participated in the event.

Due to the limited bandwidth available for data transmission from the South Pole, 
events triggering less than eight stations were prescaled by a factor of eight 
during the operation of IT59, and by a factor of three during IT73 and IT81. 
Events triggering more than eight stations were not prescaled.

%-------------------------------------
\subsection{Data Sample and Simulation}
%-------------------------------------

The prescaling scheme described above was used to divide the data into two samples: 
a ``low-energy" data set, containing events with at least three but less than eight 
stations triggered, and a ``high-energy" data set that contains events where eight 
or more stations were triggered.

During the operation of IT59, IT73, and the first year of IT81, a total of 
$3.55 \times 10^8$ events with more than 3 triggered stations were recorded.  Of 
these events, $2.90 \times 10^8$ were classified as ``low-energy" events while the 
``high-energy" sample contains the remaining $0.65 \times 10^8$ events.  A zenith 
angle cut (described below) was used to remove misreconstructed events at large 
zenith angles.  This cut reduced the final sample to $2.86 \times 10^8$ events
in the low-energy set and $0.64 \times 10^8$ events in the high-energy sample.

The angular resolution of the shower reconstruction algorithm and the median energy of 
the data sets were determined using simulated cosmic ray air showers.  Events were
generated with the CORSIKA Monte Carlo code \citep{Heck:1998a} and passed through a full 
simulation of the IceTop detector \citep{Abbasi:2012nn}.  The median energy of the samples 
determined using this simulation will depend on the assumptions made about the chemical 
composition of the primary cosmic rays. The detailed primary composition has not been directly 
measured for energies beyond 100\,TeV, but models that extrapolate existing measurements to higher energies 
indicate that in the energy range of this analysis, the cosmic ray flux consists mainly of
protons, helium, and iron \citep{Gaisser:2011cc}.  Their relative contribution is a function 
of energy, with helium and protons dominating around 100\,TeV and iron becoming the dominant 
element above several tens of PeV.  Given the uncertainties in the composition, we have 
generated only proton and iron showers as the two limiting cases for the chemical composition.  
The true median energy of the sample should be contained in the interval defined by these two 
cases.

\begin{figure}[t]
  \begin{center}	
    \includegraphics[width=0.4\textwidth]{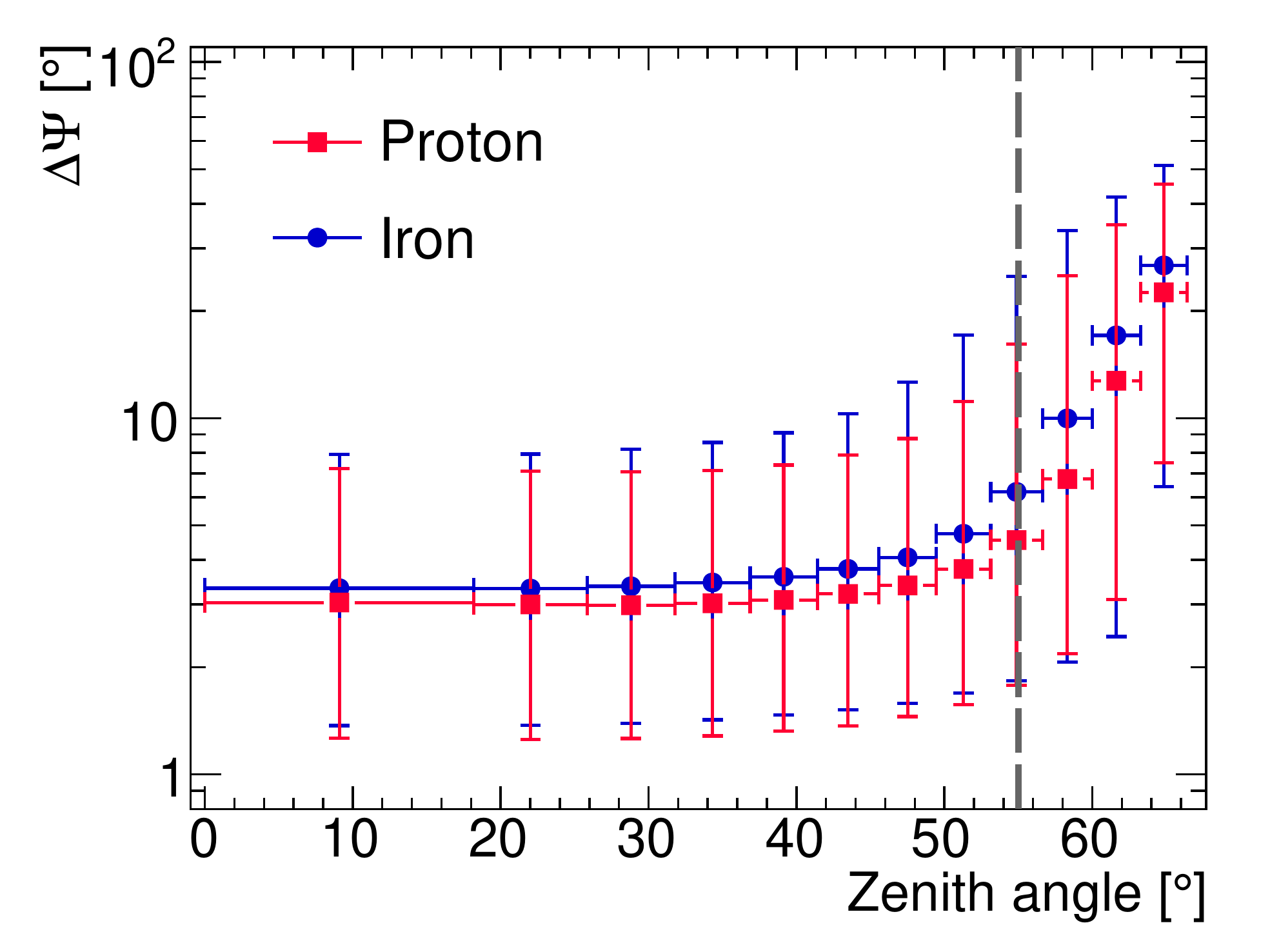}
    \caption{Median opening angle $\Delta\Psi$ between reconstructed and true arrival
             direction as a function of reconstructed zenith angle $\theta$.  At large zenith 
             angles the fraction of misreconstructed events increases.  For this reason, a 
             zenith cut was implemented that restricts the analysis to events with 
             $\theta < 55^{\circ}$ (dashed gray line).  The error bars correspond to a 68\% 
             containing interval.  IT59, IT73, and IT81 show the same dependence of angular 
             resolution on reconstructed zenith angle.}
    \label{fig:angres}
  \end{center}
\end{figure}

The arrival direction of cosmic ray showers in IceTop is reconstructed by fitting a plane
to the front of the air shower.  The fit algorithm implements an analytic $\chi^2$-minimization 
that uses the positions and hit times of the triggered stations to reconstruct the direction 
vector of the shower.  From simulation we have determined the median angular resolution 
of this algorithm to be $3^{\circ}$ for both proton and iron showers for all detector 
configurations.  In other IceTop analyses, a more precise reconstruction algorithm is used 
that takes into account the curvature of the air shower front and can reach a sub-degree resolution. 
The plane fit is better suited to our needs since it provides a resolution that is several times smaller 
than the typical angular scale of the anisotropic pattern ($> 20^{\circ}$) without requiring a larger 
number of stations triggered which would reduce the size of the cosmic ray sample.
As shown in Fig.~\ref{fig:angres}, the resolution of the plane fit degrades rapidly for 
showers with zenith angles larger than $60^{\circ}$.  For this reason, this analysis is 
limited to events with a reconstructed zenith angle smaller than $55^{\circ}$.

\begin{figure}[t]
  \begin{center}	
  \includegraphics[width=.4\textwidth]{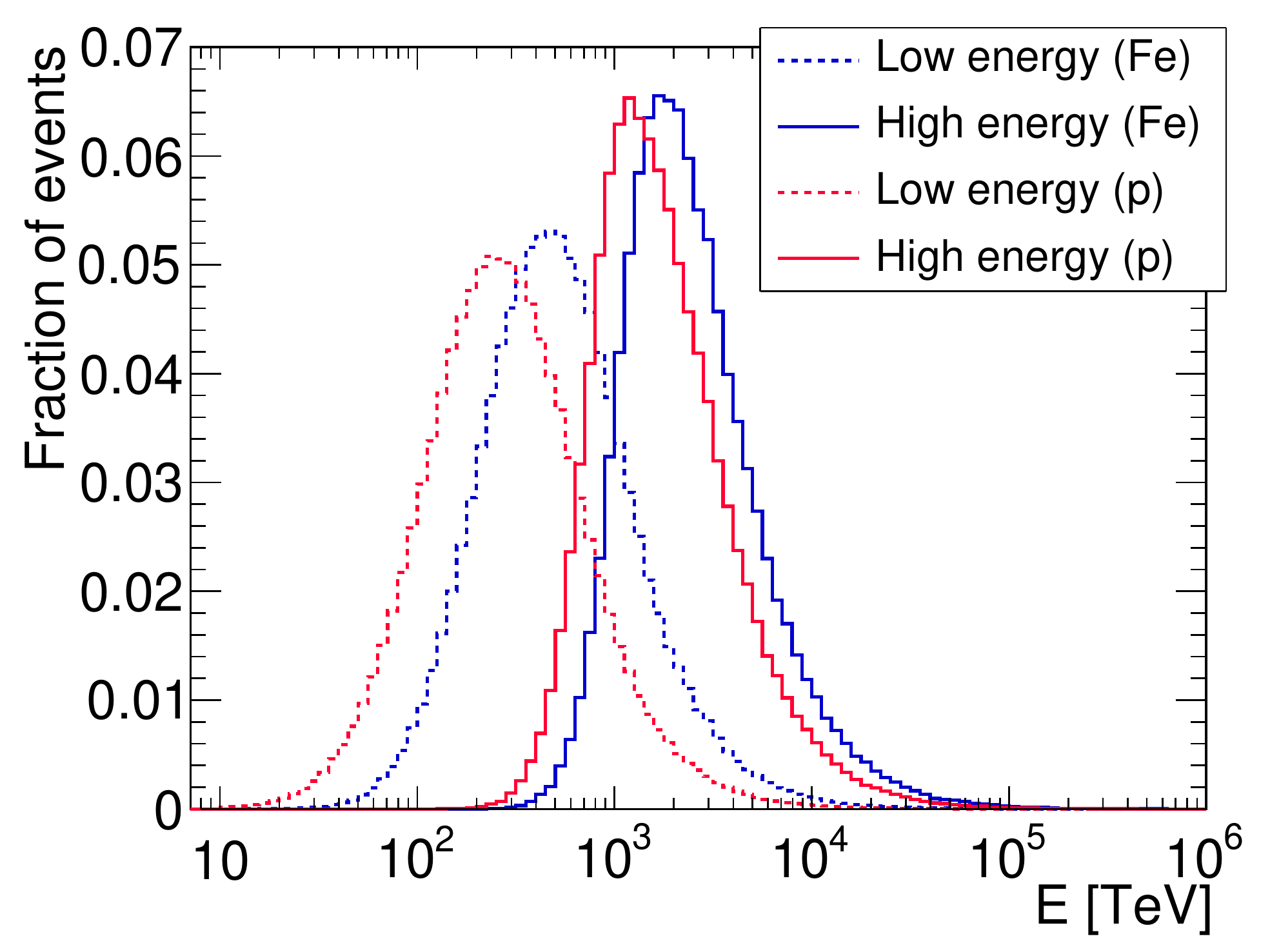} 
   \end{center} %
    \caption{Simulated energy distributions for all events in the low-energy (\emph{dashed}) and 
    high-energy (\emph{solid}) data sets assuming all-iron (\emph{blue}) and all-proton (\emph{red}) 
    compositions.  The energy distributions are the same for the IT59, IT73 and IT81 configurations.}
    \label{fig:edistro}
\end{figure}

The median energies of the data sets were determined from the energy distribution of the 
simulated air showers which satisfy the same trigger conditions as events in the
low- and high-energy data samples.  The simulated energy distributions are shown
in Fig.\ref{fig:edistro}.  The median energies and the 68\% containing intervals for the two 
composition models for each energy band are shown in Table~\ref{tab:energies}, which shows 
that the low-energy band has a median energy in the range 270 - 500 TeV, while the median energy 
for the high-energy sample should be contained in the range 1.6 - 2.2 PeV.

\begin{table*}
    \centering
    \begin{tabular}{lcccc}
    \toprule
    & \multicolumn{2}{c}{{\bf Low energy}} & \multicolumn{2}{c}{{\bf High energy}} \\
    \cmidrule(r){2-3}
     \cmidrule(r){4-5}
    Composition         & $\tilde{E}$ & 68\% interval & $\tilde{E}$  & 68\% interval  \\
    \midrule                                          
    Proton      & 0.27 PeV & 0.11- 0.69 PeV  & 1.6 PeV & 0.83 - 3.8 PeV  \\
    Iron     & 0.50 PeV & 0.22 - 1.2 PeV  & 2.2 PeV & 1.2 - 5.3 PeV    \\
    \bottomrule
    \end{tabular}
    \caption{\label{tab:energies} Median energy and 68\% containing interval in PeV for the two energy bands used in this work assuming that the 
    cosmic rays consists of either protons or iron nuclei.}
\end{table*}

\begin{figure*}[t]
  \begin{center}	
  $\begin{array}{cc} %
  \includegraphics[width=.4\textwidth]{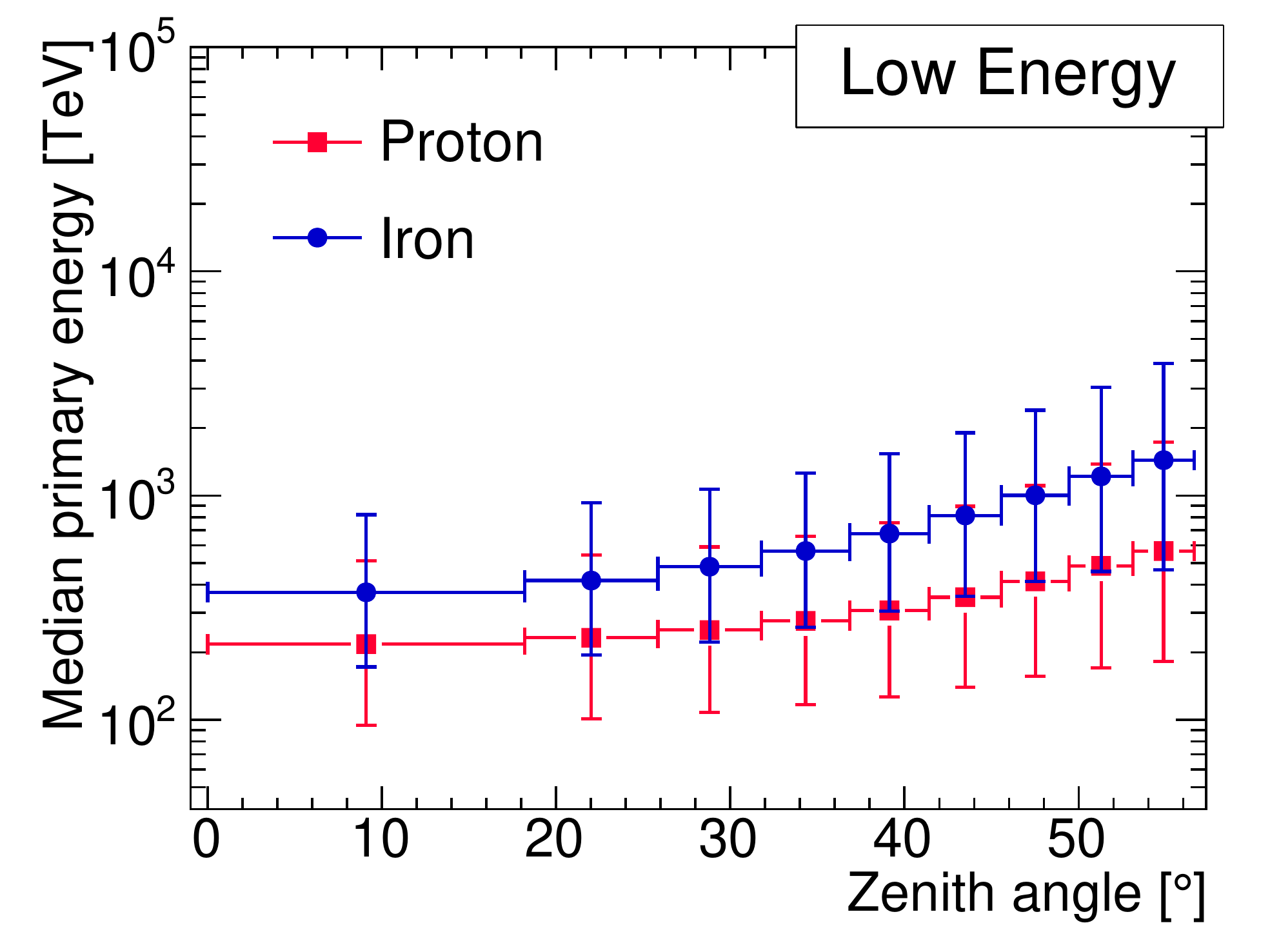} & %
   \includegraphics[width=.4\textwidth]{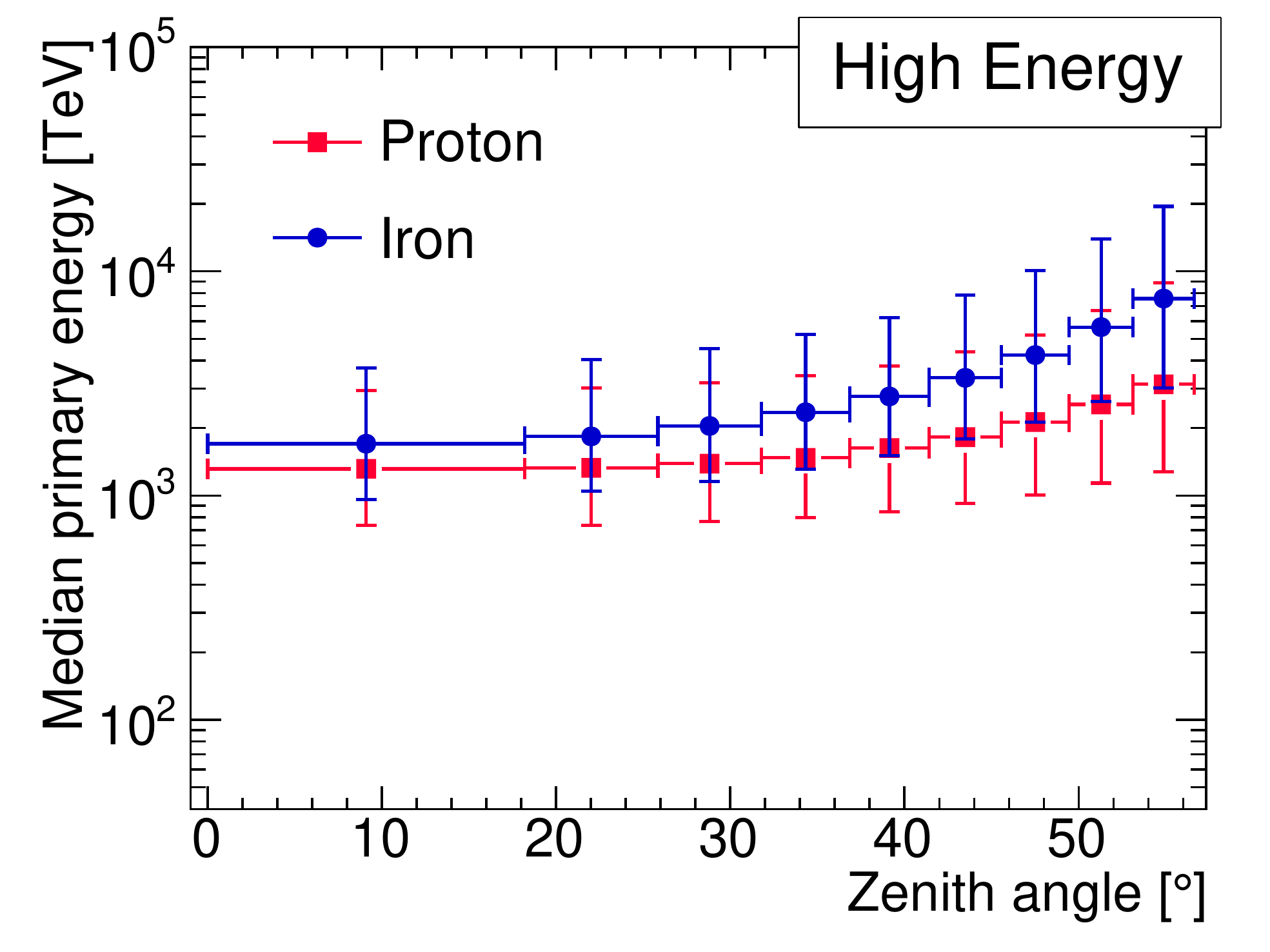} %
   \end{array}$ %
   \end{center} %
    \caption{Median energy as a function of reconstructed zenith angle for the low-energy \emph{(left)}, and 
    high-energy \emph{(right)} data sets for proton and iron cosmic ray primaries. The error bars correspond 
    to a 68\% containing interval.  IT59, IT73 and IT81 show the same energy dependence on the reconstructed 
    zenith angle.} 
    \label{fig:ezen}
\end{figure*}

For both data samples, the median energy of the primary cosmic rays monotonically increases 
with zenith-angle, as illustrated in Fig.~\ref{fig:ezen}. 

%************************************************************************************
\section{Data Analysis and Results}\label{sec:Analysis}
%************************************************************************************

%-------------------------------------
\subsection{Map Making Procedure}
%-------------------------------------

The search for anisotropy in the IceTop data is based on techniques applied in 
the analysis of cosmic ray data with IceCube and described in more detail in 
\citet{Abbasi:2011ai}.  To obtain a skymap of the relative intensity of the 
cosmic ray flux, the map of reconstructed arrival directions in equatorial 
coordinates is compared to an estimate of the isotropic expectation represented 
by a ``reference map."  The reference map is constructed from the data using 
the time-scrambling method \citep{Alexandreas:1993}.   For each detected event, 
20 ``fake" events are generated by keeping the local zenith and azimuth angles 
fixed and calculating new values for equatorial coordinates using event times 
randomly selected from within a time window $\Delta t$ bracketing the event.  The
fake events are stored in the reference map with a weight of $1/20$.  In order 
to be sensitive to anisotropy at all angular scales, we use $\Delta t = 24$\,hr.
The stability of the detector over this time was verified by performing a 
$\chi^2$-test where the distributions of zenith and azimuth coordinates for 
the events are compared inside the window. 

By using the time-scrambling algorithm, the events included in the reference map 
have the same arrival direction distribution as the data in local coordinates.  In addition, the 
method compensates for variations in the event rate, including gaps in the detector 
uptime.  We note, however, that the method is known to create artificial deficits 
near strong excesses, and vice versa near large deficits.  With a 24 hour scrambling
window, a single strong excess (deficit) will raise (lower) the reference map level 
for the entire right ascension band at the declination of the excess (deficit) which could 
bias the observed amplitude of the anisotropy and its statistical significance. This effect can become 
important if extended regions of strong excess or deficit flux are present 
in the data, and the resulting skymaps should be interpreted carefully with this
potential bias in mind.

\begin{figure}[t]
  \begin{center}	
  $\begin{array}{c} %
  \includegraphics[width=.45\textwidth]{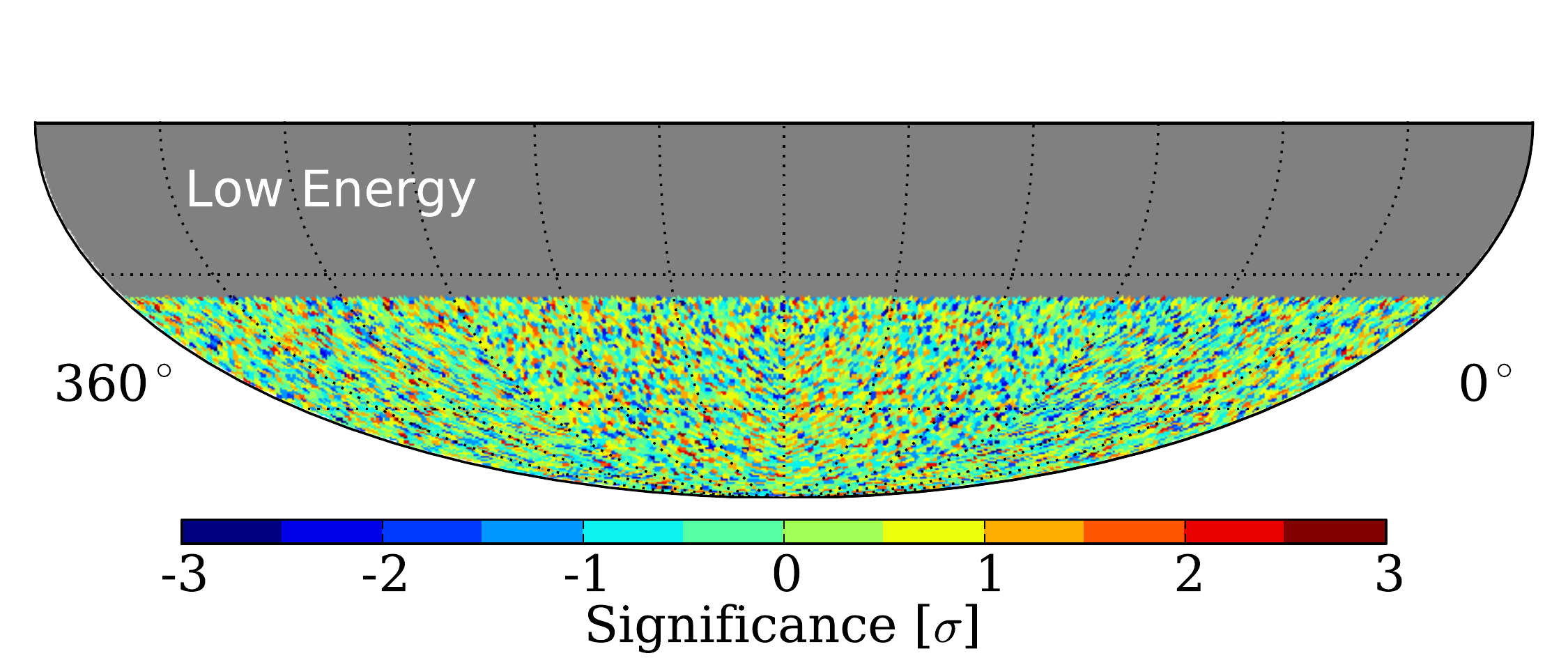} \\ %
   \includegraphics[width=.45\textwidth]{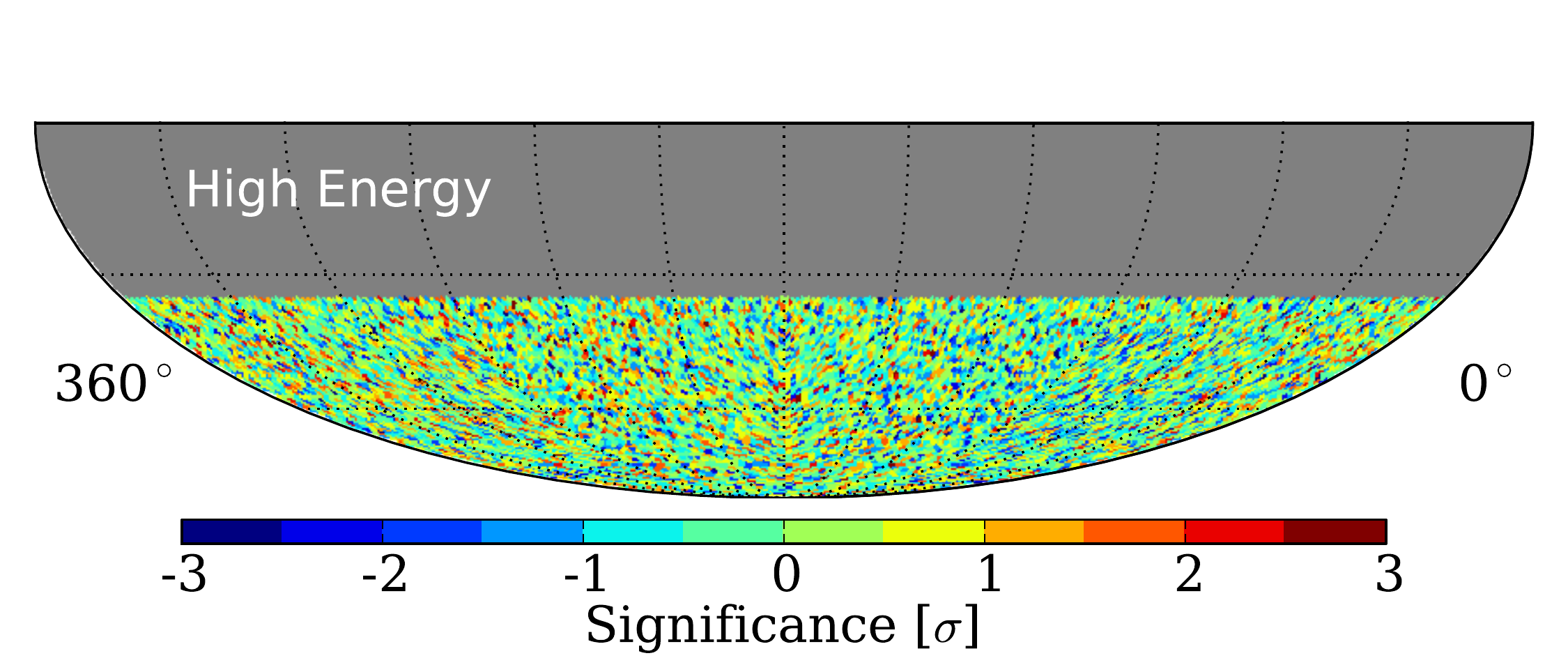}
   \end{array}$ %
   \end{center} %
    \caption{Maps of statistical significance for the low-energy (\emph{top}) and high-energy (\emph{bottom}) data sets
    with no smoothing applied.}
     \label{fig:nosmooth}
\end{figure}

The maps are built using the HEALPix equal-area pixelisation of the sphere 
\citep{Gorski:2004by} with an average pixel size of about $0.9^{\circ}$. Maps of statistical 
significance for the low- and high-energy data sets are shown in Fig.~\ref{fig:nosmooth}.
The bin size is not optimized for a study of anisotropy at scales larger than 
the angular resolution of the detector.  We therefore apply a smoothing procedure 
to both the data map and the reference map in order to increase the sensitivity 
of the method to structures with larger angular sizes.  The smoothing procedure 
is essentially a rebinning of the maps, but rather than producing maps with
fewer (but statistically independent) pixels, we retain the original $0.9^{\circ}$ 
binning.  At each bin, we add the counts from all pixels within some angular radius 
(``smoothing radius'') of the bin.  This produces maps with Poisson uncertainties, 
but with bins that are not statistically independent.  Since the optimal smoothing scale
is not known {\it a priori}, we perform a scan over different smoothing angles. 
After smoothing, the relative 
intensity $\delta I_i$, {\it i.e.} the amplitude of deviations from the isotropic
expectation for each angular bin $i$, is calculated using:
\eq{
	\delta I_i = \frac{\Delta N_i}{\langle N \rangle_i} 
            = \frac{N_i - \langle N \rangle_i}{\langle N \rangle_i} ~~,
}
where $N_i$ and $\langle N \rangle_i$ are the number of events in pixel $i$ of the 
data map and the reference map, respectively.  The statistical significance of the 
observed deviations is calculated using the method described in \citet{LiMa:1983}. 

The significance is later corrected to account for the number of trials introduced 
by looking everywhere in the sky for significant fluctuations, and for the fact that 
a scan was performed over different smoothing radii to search for anisotropy at 
different angular scales.

%-------------------------------------
\subsection{Results}
%-------------------------------------

\begin{figure*}[t]
  \begin{center}	
  $\begin{array}{cc} %
  \includegraphics[width=.5\textwidth]{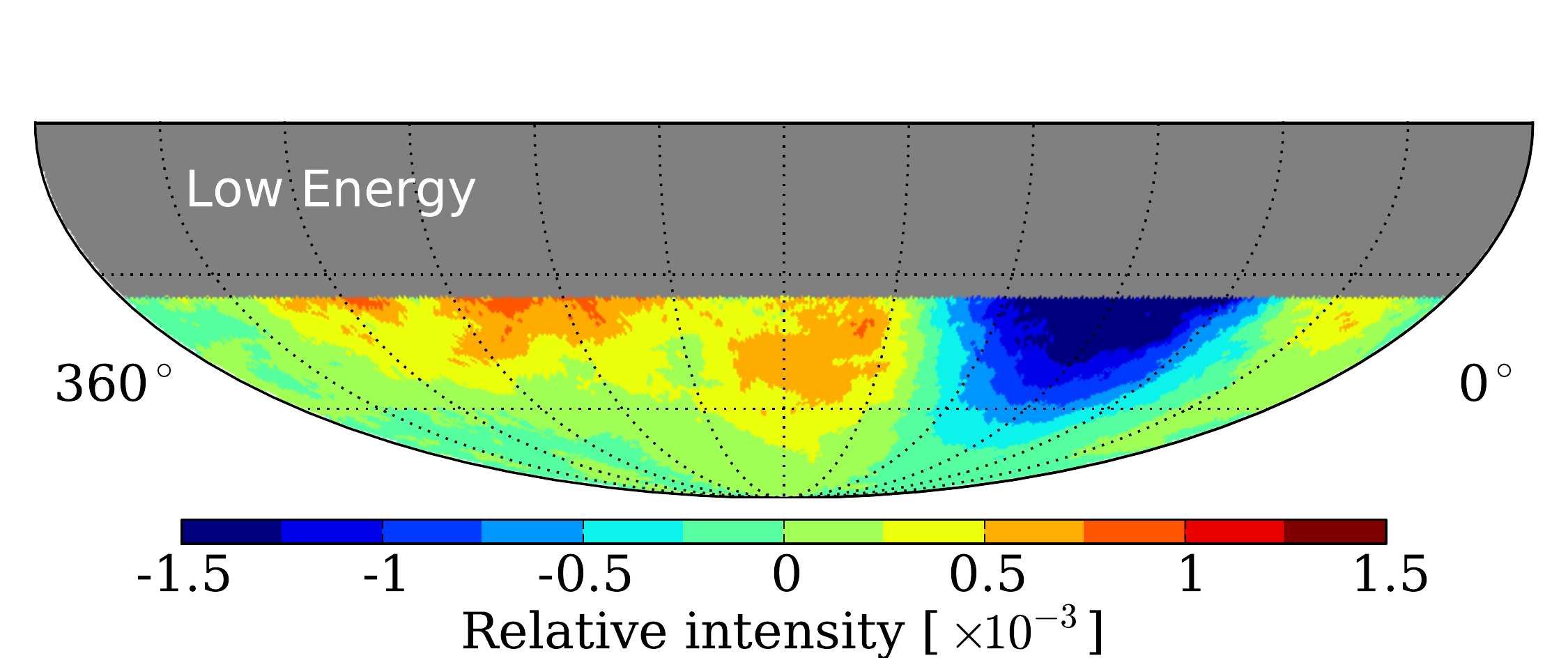} & %
   \includegraphics[width=.5\textwidth]{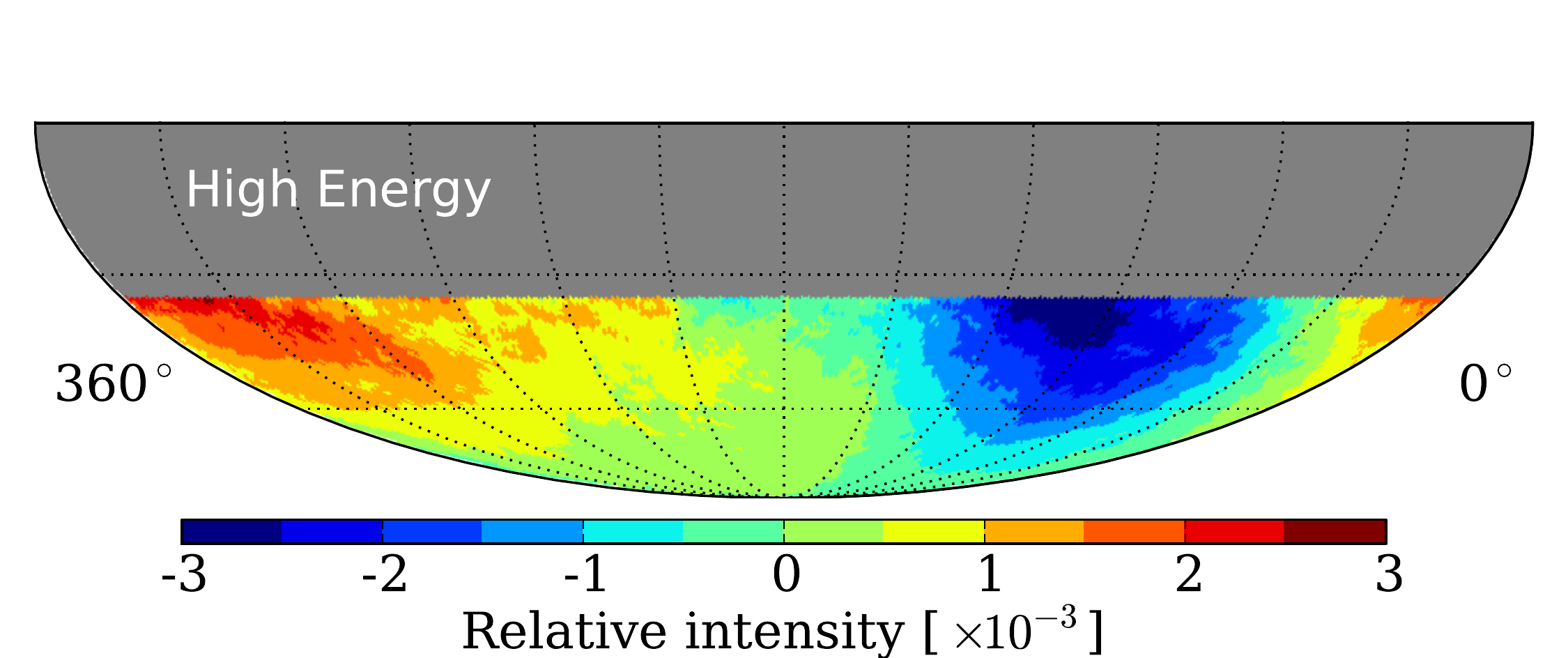} \\ %
   \includegraphics[width=.5\textwidth]{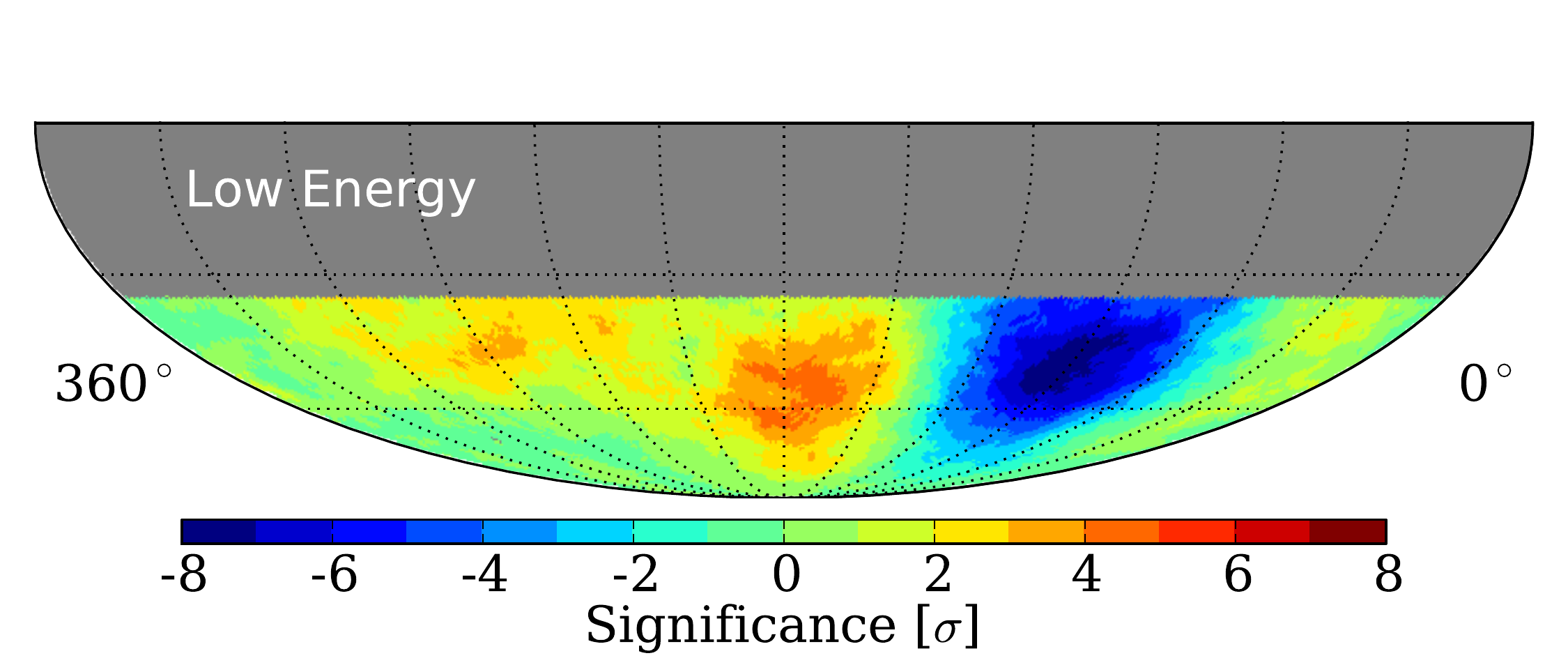} & %
   \includegraphics[width=.5\textwidth]{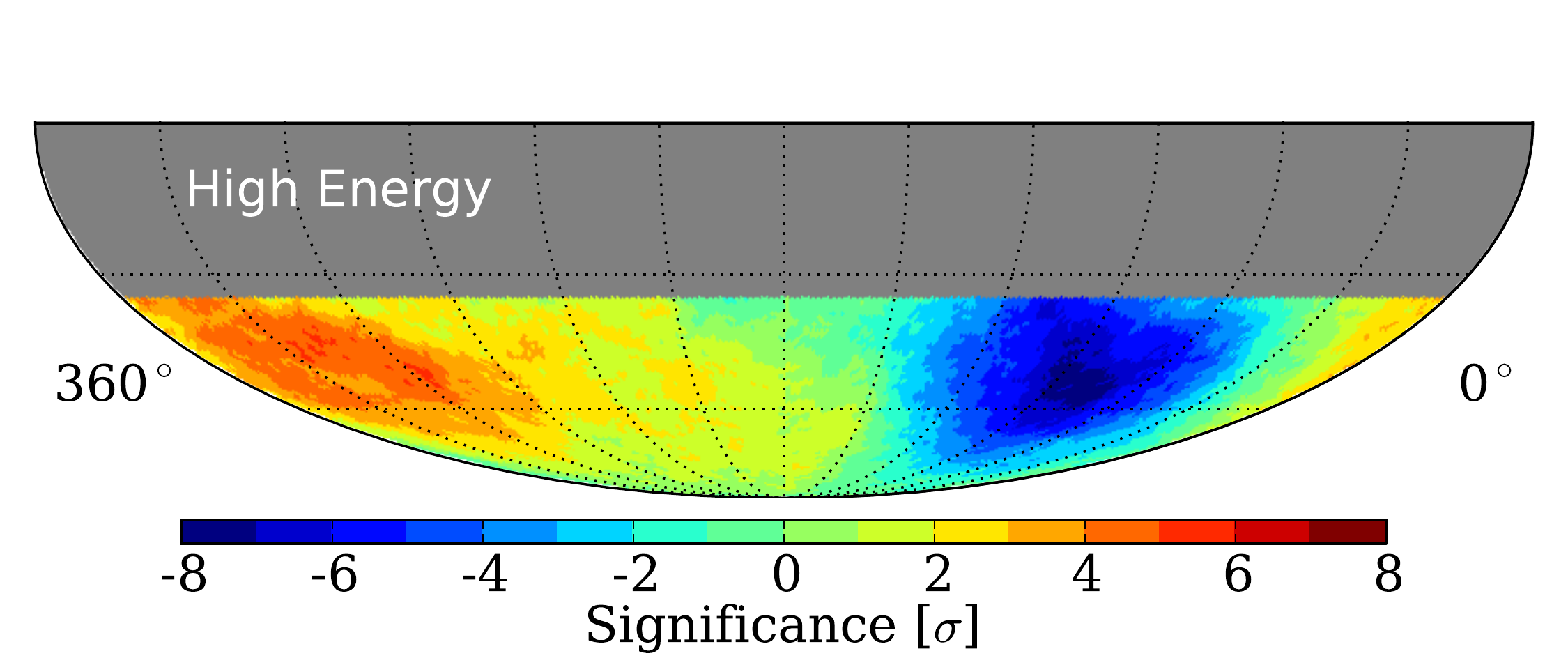} %
   \end{array}$ %
   \end{center} %
    \caption{Relative intensity (\emph{top}) and statistical significance (\emph{bottom}) maps for the low-energy
     (\emph{left}) and high-energy (\emph{right}) data sets for a smoothing angle of $20^{\circ}$.}
     \label{fig:maps}
\end{figure*}

The smoothing procedure described above was applied to the low- and 
high-energy maps for smoothing radii between $5^{\circ}$ and $50^{\circ}$ in $3^{\circ}$
steps. A search for regions of high significance was performed on the resulting smoothed
maps.  The relative intensity and significance maps for the low- and high-energy data 
are shown in Fig.~\ref{fig:maps} for a representative smoothing radius of $20^{\circ}$ where
all the relevant features observed in these two energy ranges are visible. Maps for all 
other smoothing radii are available on the web as supplemental information to this paper.

The low-energy map is dominated by a strong deficit in relative intensity.  The statistical 
significance of the deficit reaches a maximum of $8.5\sigma$ for a smoothing radius of 
$29^{\circ}$ at a location around ($\alpha=85.8^{\circ}$, $\delta = -36.4^{\circ}$).  Since 
the search for this minimum is performed over about 10000 pixels in the map, and across all 
16 different smoothing radii, there is a trials factor of at most $1.6 \times 10^{5}$ that 
reduces the post-trial significance of the deficit to $7.0\sigma$.  It must be noted that 
this correction for trials is conservative, since the pixels in the map are statistically 
correlated by the smoothing procedure, which results in a smaller effective number of trials 
than the maximum.

For the optimal smoothing radius of $29^{\circ}$, the relative intensity $\delta I$ reaches a value of about
$-1.5 \times 10^{-3}$ at the location of the greatest deficit around ($\alpha = 83.7^{\circ}$, 
$\delta = -35.7^{\circ}$), near the edge of our exposure window.  Differences in declination 
between the location of the maximum relative intensity and maximum significance are due to 
the fact that the statistical significance accounts for both signal strength and the 
declination-dependence of our statistics.  This usually implies that the position of maximum 
significance is offset towards lower declination values where the statistics increase. 

Also visible in the low-energy map is a region of excess flux located around 
($\alpha = 182.9^{\circ}$, $\delta=-55.9^{\circ}$).  The maximum pre-trial significance for 
this region is $5.3\sigma$ for a smoothing angle of $26^{\circ}$.  The significance falls 
below the $3\sigma$ threshold after accounting for trials. 

As mentioned above, in the presence of a strong deficit the time-scrambling algorithm can 
introduce an underestimation of the isotropic reference level, which can produce 
spurious excess regions in the parts of the sky surrounding the signal region.  For this 
reason, it is possible that the excess observed in the low-energy data set is associated 
with the presence of the deficit region.

\begin{figure*}[t]
  \begin{center}
    $\begin{array}{cc} %
  \includegraphics[width=.45\textwidth]{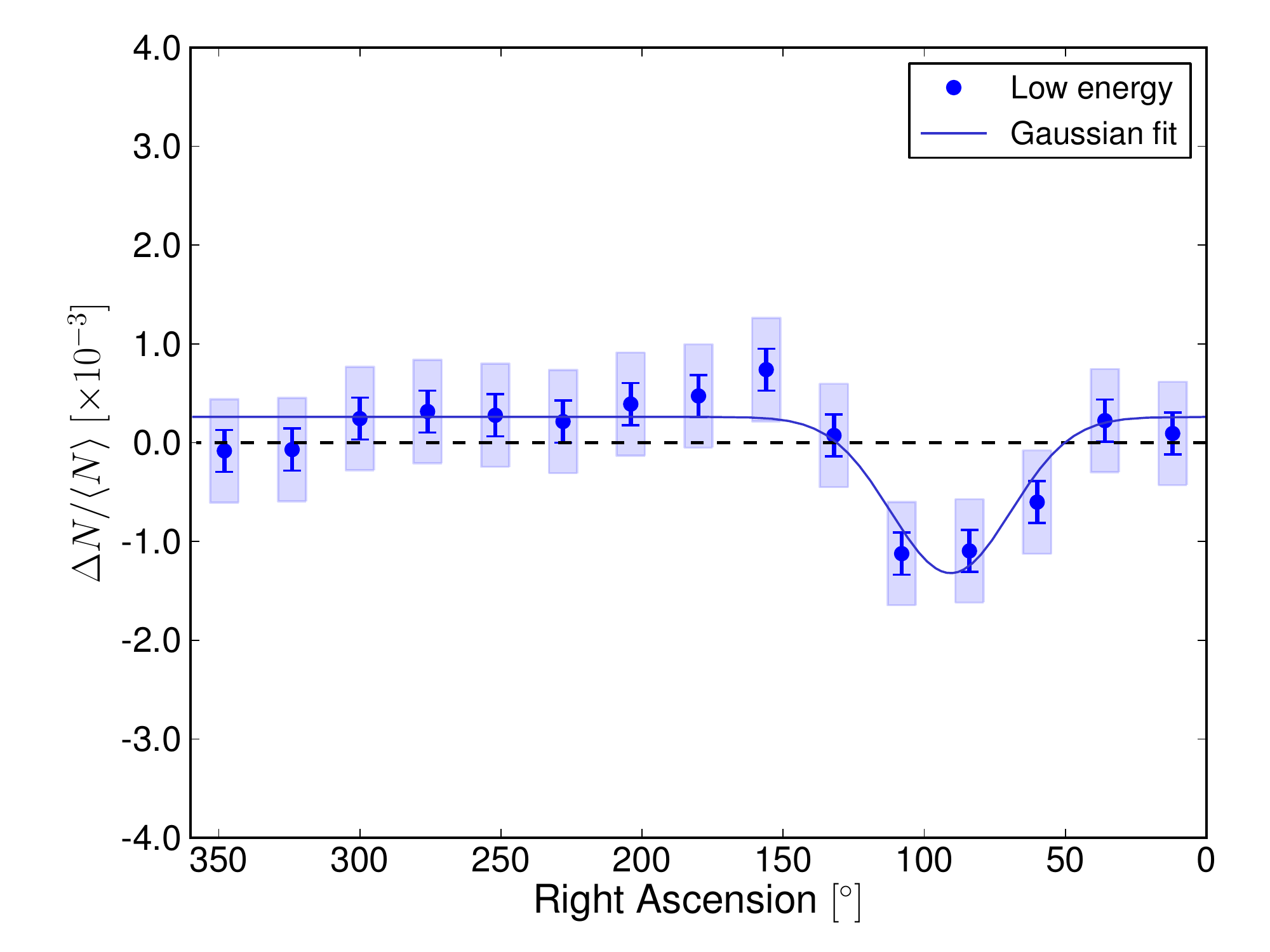} & %
   \includegraphics[width=.45\textwidth]{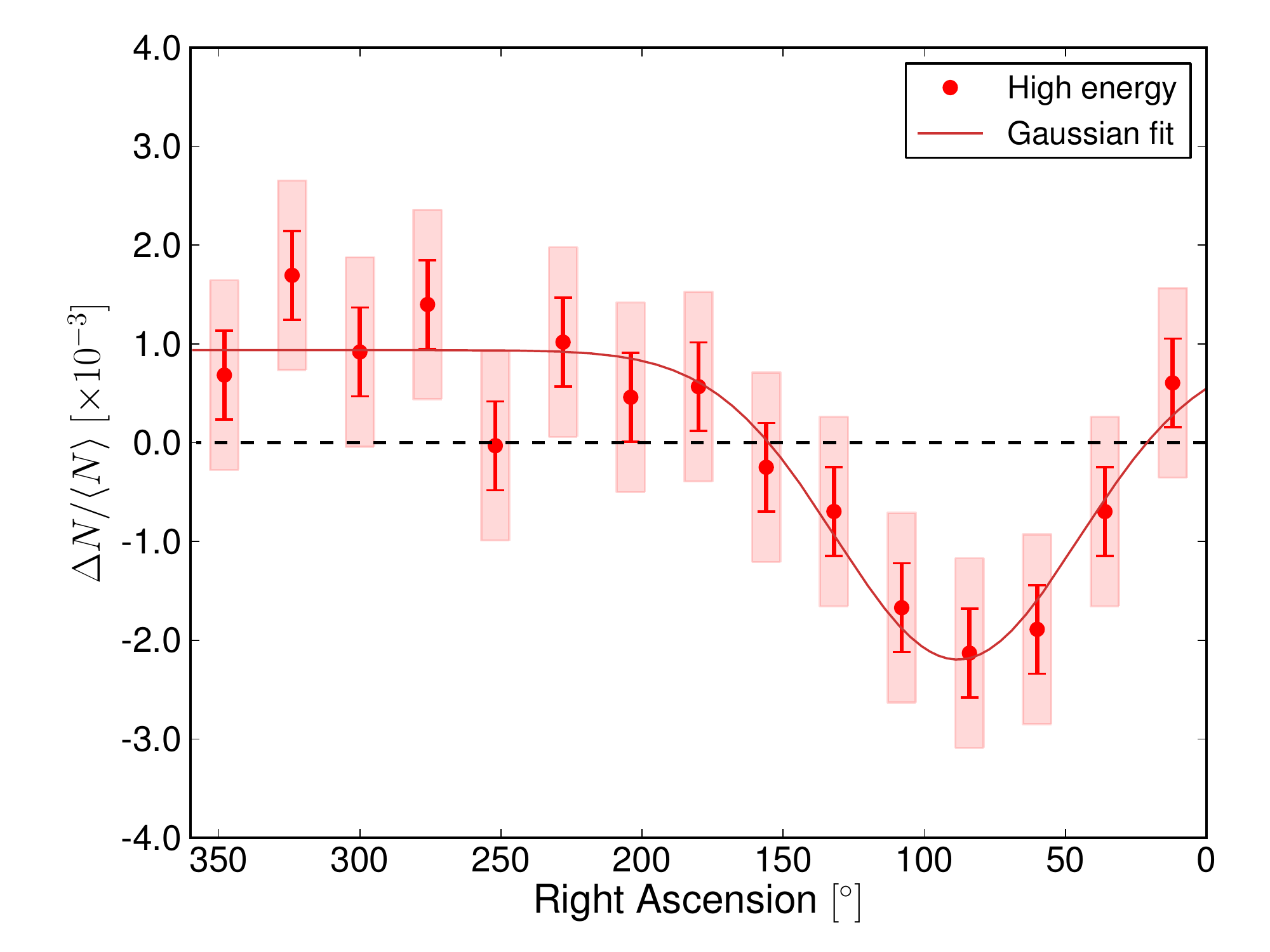} \\ %
   \end{array}$ %
   \end{center} %
    \caption{Relative intensity as a function of right ascension for the low-energy (\emph{left}) and 
    high-energy (\emph{right}) data samples in the declination band $-75^{\circ} < \delta < -35^{\circ}$. 
    The error bars are statistical while the colored boxes indicate the systematic uncertainty obtained 
    from analyzing the same data in the anti-sidereal time frame (see Section~\ref{sec:systematics} for 
    details). The result of a fit using the Gaussian function given in Eq.~\ref{eq:gauss} to both energy 
    bands are also shown.}
    \label{fig:raproj}
\end{figure*}

The high-energy map also shows statistically significant anisotropy which is dominated by 
a deficit located in the same approximate position as that observed in the low-energy data. 
The pre-trial significance of the deficit is $8.6\sigma$ ($7.1\sigma$ post-trials) for a 
smoothing angle of $35^\circ$, with its minimum located at ($\alpha = 79.4^{\circ}$,
$\delta=-37.2^{\circ}$).  The main difference between the low- and high-energy deficits 
is that the value of $\delta I$ for the greatest deficit in the high-energy sample is $-2.3 \times 10^{-3}$, larger 
than its low-energy counterpart.  This is evident in Fig.~\ref{fig:raproj}, where the 
relative intensity is projected onto the right ascension axis using the declination band 
$-75^{\circ} < \delta < -35^{\circ}$. 

A second notable feature in the high-energy map is a wide excess region that reaches a peak 
significance of $5.9\sigma$ ($3.4\sigma$ post-trials) for a smoothing angle of $41^{\circ}$.  
The excess does not appear to be concentrated in any particular part of the sky, but distributed 
across a wide band in right ascension.  This is visible in the one-dimensional projection shown 
in Fig.~\ref{fig:raproj}, where the relative intensity reaches a plateau above $\alpha > 170^{\circ}$ 
which is offset from zero by about $10^{-3}$.  As in the low-energy data set, such an excess 
could be associated with the presence of the observed deficit in the same declination band 
that introduces a bias in the reference-level estimation.

In order to characterize the observed anisotropic pattern, we attempted to fit the relative 
intensity projections of the data shown in Fig.~\ref{fig:raproj} with the first terms 
(dipole and quadrupole) of a harmonic series:
\eq{
  \delta I(\alpha) = \sum_{j=1}^{2} A_j  \cos[j(\alpha - \phi_j)]+ B ~~.
  \label{eq:dq}
}
The harmonic fit parameters for the low-energy dataset are $A_1 = (5.53 \pm 0.91) \times 10^{-4}$,
$\phi_1 = -111.7^{\circ} \pm 9.7^{\circ}$, $A_2 = (4.03 \pm 0.80) \times 10^{-4}$, $\phi_2 = 1.0^{\circ} \pm 7.9^{\circ}$,
and $B = (-0.47 \pm 0.66) \times 10^{-4}$. For all parameters the indicated uncertainties are statistical. 
The $\chi^2$/dof for the fit is 18.4/10.

In the case of the high-energy dataset the fit parameters are $A_1 = (1.43 \pm 0.19) \times 10^{-3}$,
$\phi_1 = -84.7^{\circ} \pm 8.0^{\circ}$, $A_2 = (7.34 \pm 1.9) \times 10^{-4}$, $\phi_2 = -14.7^{\circ} \pm 7.7^{\circ}$, and
$B = (-0.14 \pm 1.35) \times 10^{-4}$. In this case, the $\chi^2$/dof for the fit is 6.3/10.

In the case of the low-energy dataset, the reduced $\chi^2$ of the fit is considerably larger than unity,
indicating that for this data, as in the case of the previous observation of anisotropy at 400\,TeV with IceCube \citep{Abbasi:2011zka}, 
this choice of harmonic base functions does not fit the data particularly well. For this reason, a new fit is performed using the following 
Gaussian function: 
\eq{
	\delta I(\alpha) = A e^{-(\frac{\alpha - \alpha_s}{\sqrt{2} \sigma})^2} + b ~~,
\label{eq:gauss}
}
where $\alpha$ is right ascension, $A$ is the amplitude, $\sigma$ is the width, and $\alpha_s$ is 
the right ascension of the center of the deficit.  The parameter $b$ represents an overall offset 
from isotropy that can be introduced by the presence of a strong signal in the data.

The results of these fits are shown in Table~\ref{tab:fits}, and indicate that while the center 
point of the deficit for both data sets is consistently located at $\alpha_s \sim 90^{\circ}$, 
both the amplitude and the width are larger in the high-energy sample, with both values increasing 
by about a factor of two with respect to the low-energy case. The location of the deficit in the right
ascension projection (Fig.~\ref{fig:raproj}) is consistent with its location in the skymap (Fig.~\ref{fig:maps}), 
within statistical and systematic uncertainties, for both the low-energy and high-energy samples. Similarly, the amplitudes
in relative intensity of the deficit agree well, when the overall offset $b$ is taken into account.

\begin{table*}
    \centering
    \begin{tabular}{lll}
    \toprule
    & {\bf Low energy} & {\bf High energy} \\ 
    \midrule                                          
    $A$ & $ (-1.58 \pm 0.46 \pm 0.52) \times 10^{-3} $  & $ (-3.11 \pm 0.38 \pm 0.96) \times 10^{-3} $ \\
$\alpha_s$ & $90.6^{\circ} \pm 6.8^{\circ} \pm 9.3^{\circ}$ & $88.1^{\circ} \pm 6.8^{\circ} \pm 11.1^{\circ}$ 	\\
$\sigma$ & $21.3^{\circ} \pm 5.8^{\circ} \pm 7.6^{\circ}$ & $43.1^{\circ} \pm 7.3^{\circ} \pm 13.1^{\circ}$	\\
$b$ & $(2.61 \pm 0.64 \pm 5.20) \times 10^{-4}$ & $(9.37 \pm 1.96 \pm 9.60) \times 10^{-4}$ \\
$\chi^2$/dof & 13.2/11 & 10.7/11 \\
    \bottomrule
    \end{tabular}
    \caption{\label{tab:fits} Fit parameters obtained for both energy datasets for the Gaussian 
function given in Eq.~\ref{eq:gauss}.  In all cases, the first quoted uncertainty is statistical 
while the second one corresponds to the systematics.}
\end{table*}

The systematic uncertainty associated with each fit value was obtained from the systematic 
uncertainty of the relative intensity projection, shown as shaded boxes in Fig.~\ref{fig:raproj}. 
The systematic uncertainty of the projection was conservatively estimated as the maximum amplitude 
of the relative intensity distribution for each energy data set when analyzed in the anti-sidereal 
time frame (see Section~\ref{sec:systematics}).

A previous search for anisotropy as a function of cosmic-ray primary energy was performed using 
 muon data from the IceCube detector \citep{Abbasi:2011zka}. In that analysis, cuts were applied to the data to
select two sets of cosmic-ray events: one with a median primary energy of 20\,TeV, and a second one
with a median energy of 400\,TeV, similar to the low-energy IceTop sample. The 400\,TeV IceCube skymap 
shows a deficit region similar to the one observed in the IceTop low-energy sample. At 20\,TeV, IceCube 
observed a large-scale structure that is consistent with previous observations at these energies 
\citep{Abbasi:2010mf}\citep{Abbasi:2011ai}. The angular size and the orientation in the sky of the 20\,TeV
anisotropy is different from that observed at 400\,TeV by IceCube and IceTop.  Note that while the IceTop 
median energy was obtained by assuming two limiting cases of primary chemical composition 
(all-proton or all-iron primaries only), the IceCube median energy was obtained assuming that 
the cosmic ray flux follows the polygonato composition models \citep{Hoerandel:2002yg}, which in 
principle could lead to some differences in the actual median energy and the energy
distribution of events in both samples.

\begin{figure}[t]
  \begin{center}
  \includegraphics[width=.45\textwidth]{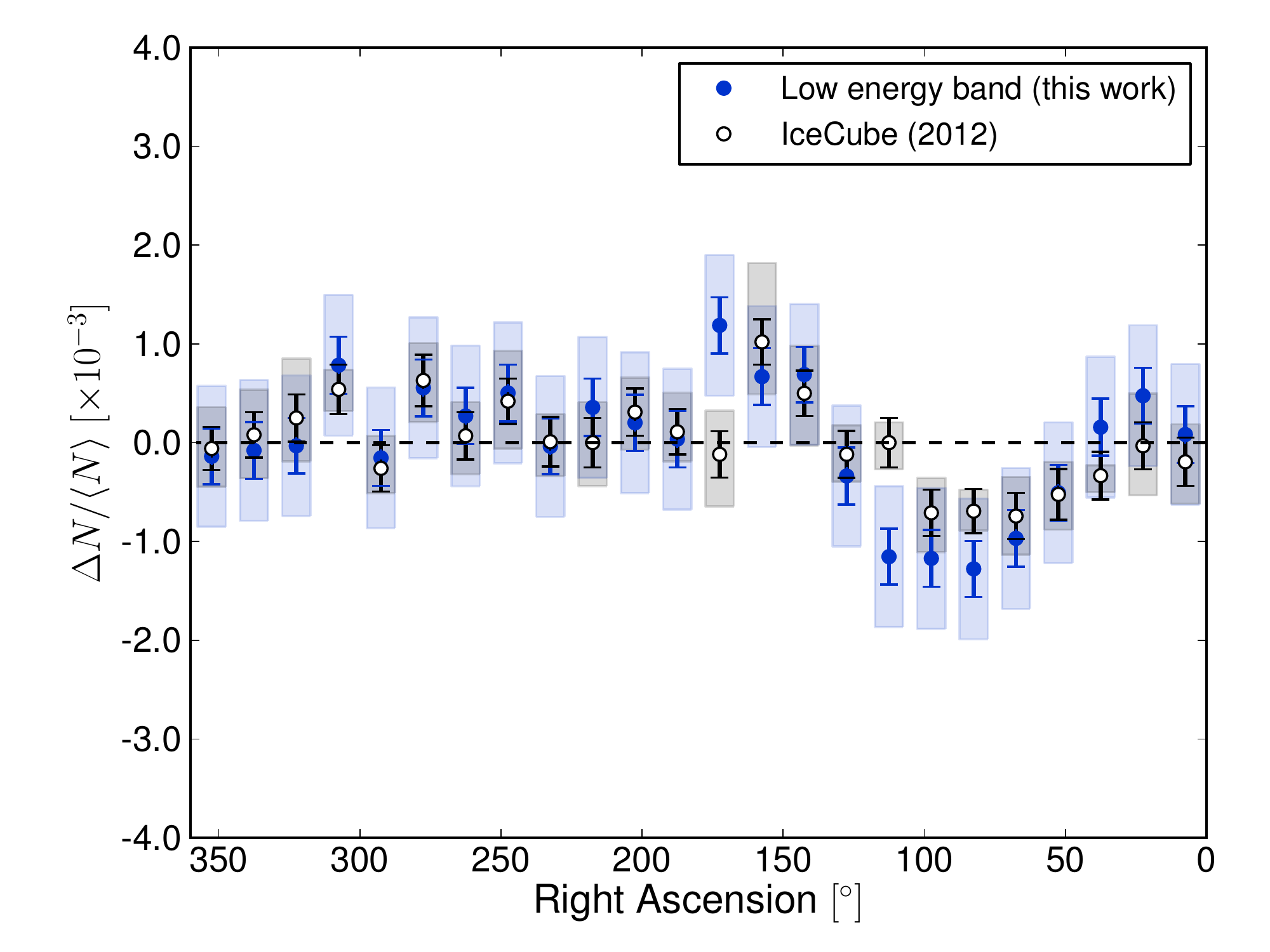}
  \end{center} %
    \caption{Comparison between the relative intensity projections for the IceTop low-energy 
    sample (\emph{blue filled circles}) and the IceCube 400 TeV sample (\emph{black open 
    circles}) reported by \cite{Abbasi:2011zka}. The location and amplitude of both deficits 
    are consistent given the statistical and systematic uncertainties. The declination range for the IceCube
    plot is $-75^{\circ} < \delta < -25^{\circ}$, slightly different from the IceTop one.}
    \label{fig:racomp}
\end{figure}

The smoothing procedure described here was also used in the IceCube analysis, and the significance 
of the deficit was maximized for a smoothing radius of $29^{\circ}$, the same as the optimal smoothing 
angle for the low-energy IceTop data.  Fig.~\ref{fig:racomp} shows a comparison between the IceCube 
and IceTop results at 400\,TeV. The location and amplitude of the deficits observed in both data sets
agree given the current values of the statistical and systematic uncertainties associated with 
both measurements.

%************************************************************************************
\section{Systematic uncertainties } \label{sec:systematics}
%************************************************************************************

A number of tests have been performed in order to quantify the systematic uncertainties associated
with the observation of anisotropy in the IceTop data. 

In the first study, the anisotropy search was performed on three independent data subsamples, 
each containing events recorded during the operation of the three different detector configurations
IT59, IT73, and IT81 considered in this work.  In this manner we can determine the possible systematic 
effect introduced by the changing geometry of the detector on the observed anisotropy. 

\begin{figure*}[t]
  \begin{center}	
  $\begin{array}{cc} %
  \includegraphics[width=.45\textwidth]{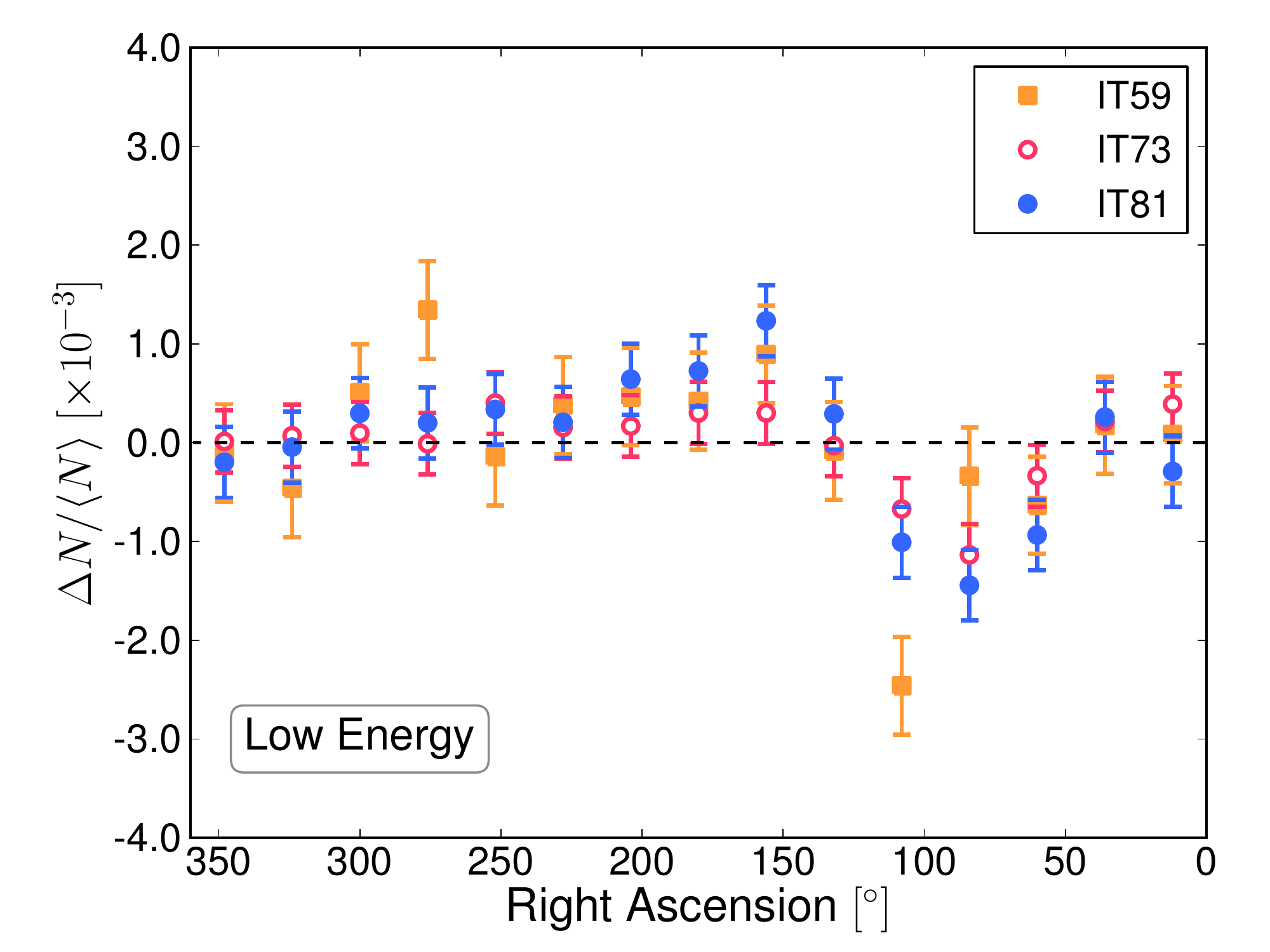} & %
   \includegraphics[width=.45\textwidth]{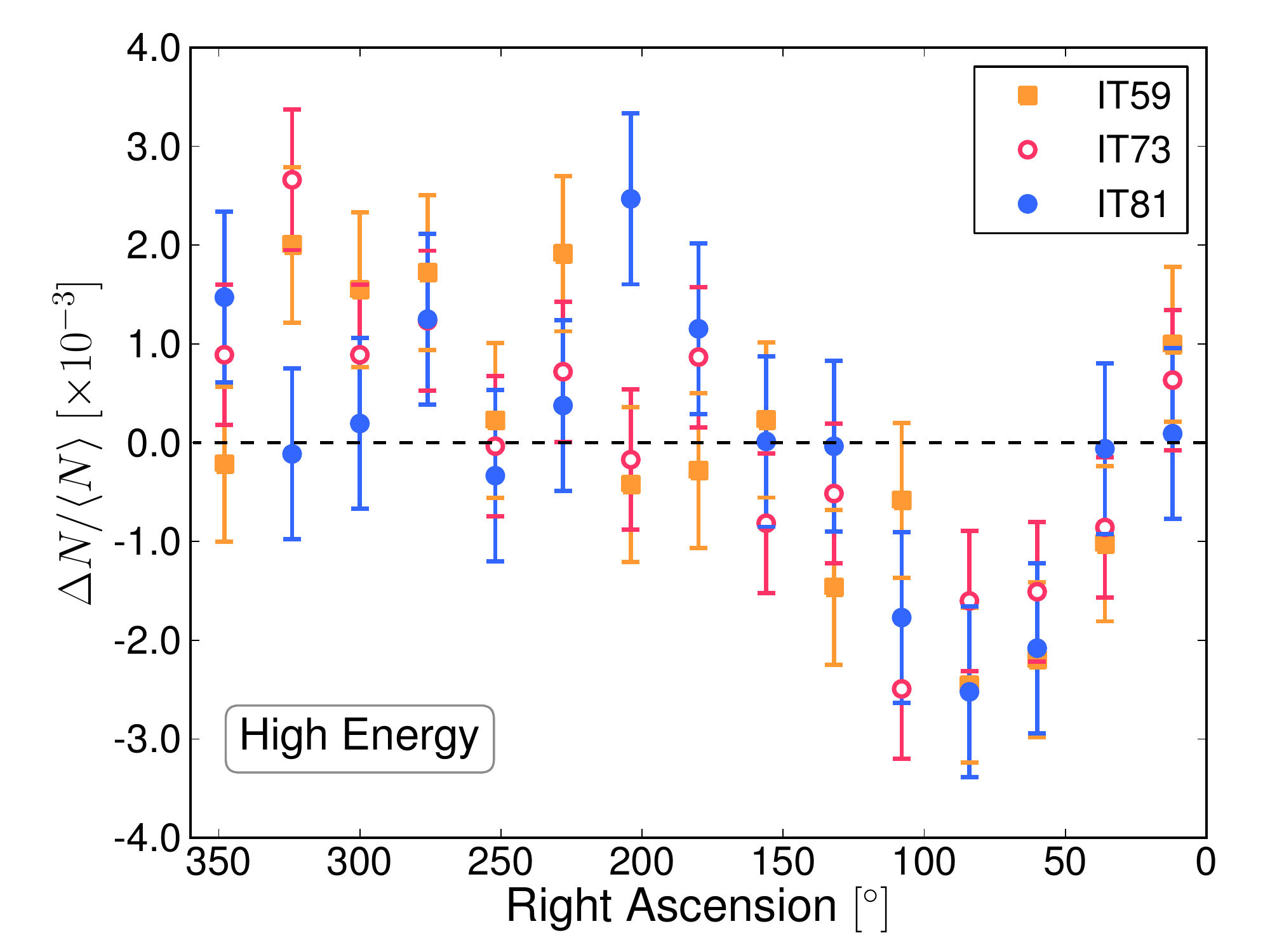}  \\ %
   \end{array}$ %
   \end{center} %
    \caption{Relative intensity as a function of right ascension for the low-energy (\emph{left}) and 
    high-energy (\emph{right}) data samples in the declination band $-75^{\circ} < \delta < -35^{\circ}$ for the three
    detector configurations of IceTop considered in this work (IT59, IT73, and IT81). For clarity, only statistical 
    error bars are shown.} 
    \label{fig:detectors}
\end{figure*}

The results of this comparison are shown in Fig.~\ref{fig:detectors}, where the relative intensity 
as a function of right ascension for the declination band $-75^{\circ} < \delta < -35^{\circ}$ is
displayed for all three detector configurations and for the low- and high-energy samples separately.  
The anisotropy observed by all three configurations is consistent within statistical uncertainties.

Another test was performed to evaluate the impact of the seasonal variation of the cosmic ray 
rate at the South Pole \citep{Tilav:2010hj}. In this study, four different time periods were 
selected from the data: June through August, September through November, December through February, 
and March through May for each year of operation of the detector.  These four data sets contain 
events taken with comparable detector geometries, but recorded during different phases of the 
seasonal variation cycle. The results of this study indicate that the anisotropy observed in each of the 
four time periods is consistent within statistical uncertainties.  

Other possible seasonal effects on the anisotropy are also investigated. First, an analysis was performed to 
look for anisotropy in the so-called ``solar'' time frame, defined as having 365.25 ({\it i.e.} 
complete revolutions in the coordinate frame) per year. The motion of the Earth around the Sun should 
create a dipolar anisotropy in the solar frame with an amplitude of $4.7 \times 10^{-4}$. No anisotropy
was observed using IceTop data. However, simulations of the solar dipole assuming the IceTop acceptance in local coordinates 
indicate that the current size of the data set is insufficient for a statistically significant observation.

The second analysis consists of a search for anisotropy analysis in the ``anti-sidereal" time frame, defined as 
having 364.25 days . No signal should be observed in this frame unless there exists a seasonal variation in the solar 
time frame that could affect the anisotropy in sidereal time (period of 366.25 days).  See \citet{Abbasi:2011ai} for details.  

We performed the anti-sidereal analysis on the combined three-year data set and obtained both 
skymaps and one-dimensional relative intensity projections for the low- and high-energy bands. 
The skymaps produced for the anti-sidereal frame do not exhibit any significant anisotropy that 
could indicate a possible systematic bias in the sidereal frame.  The systematic uncertainty of 
the sidereal anisotropy due to seasonal variations, shown in Fig.~\ref{fig:raproj},
is obtained from the relative intensity projections in the anti-sidereal frame. This uncertainty
is conservatively estimated as the maximum departure from the reference level of the anti-sidereal
right ascension distribution.

%************************************************************************************
\section{Conclusions}\label{sec:Conclusions}
%************************************************************************************

A study of cosmic ray arrival directions with IceTop at two different median
energies, 400\,TeV and 2\,PeV, shows significant anisotropy in both sets.
The skymap is dominated by a single deficit region with an angular
size of about $30^{\circ}$.  The skymap at 400\,TeV is similar to a skymap 
of comparable median energy obtained from cosmic rays in IceCube
\citep{Abbasi:2011zka}.  IceTop data show that this anisotropy persists to 2\,PeV.  

The anisotropy in the southern sky at 400\,TeV and 2\,PeV is different in shape and 
amplitude from what is observed at 20\,TeV.  In the northern hemisphere, the EAS-TOP 
experiment has also recently found indications for an increasing amplitude 
and a change of phase between 100\,TeV and 400\,TeV in a harmonic analysis in 
right ascension that considers the first and second harmonic \citep{Aglietta:2009mu}. 
The IceTop anisotropy is not well-described by a sum of a dipole and a quadrupole 
moment, so the results cannot be directly compared.  However, both northern
and southern hemisphere data seem to show qualitatively similar trends.

Although these results do not provide conclusive evidence for any particular model,
they lend support to scenarios where the large-scale anisotropy is a superposition 
of the flux from a few nearby sources.  The sparse spatial distribution and the different ages 
of nearby supernova remnants
are expected to lead to a bumpy structure in the amplitude and sudden changes in the
phase of the anisotropy as a function of energy \citep{Blasi:2011fm}.  Unfortunately, 
this energy dependence is dominated by details such as the geometry of the Galaxy, 
the location, age and injection spectrum of the sources, and the energy dependence 
of the cosmic ray diffusion coefficient.  While the predicted strength of the 
amplitude has the correct order of magnitude, further quantitative predictions are 
not possible at this point.  In addition, in their simplest form, these models 
predict a dipolar anisotropy, whereas in most cases, the observed anisotropy cannot 
be described as a simple dipole, which also means that ``amplitude'' and ``phase'' 
are not well-defined.  

It was recently pointed out that an existing dipolar flux in addition to
cosmic ray propagation in turbulent magnetic fields close to Earth can explain 
the appearance of small-scale structure \citep{Giacinti:2011mz}.  For cosmic rays
with energies from TeV to PeV, the relevant distance scale is a few tens of
parsecs, so the observed anisotropy at these energies is indicative of the
turbulent Galactic magnetic field within this distance from Earth.  The model
predicts that the anisotropy is energy-dependent, but again, due to our poor 
knowledge of interstellar magnetic fields, it cannot provide more quantitative 
predictions that can be tested with data.  A detailed measurement of
the anisotropy might lead to a better understanding of these fields.

The observation of cosmic ray anisotropy with IceTop opens up new possibilities
for future studies that go beyond mapping the arrival direction distribution as 
a function of energy.  IceTop is designed to measure the energy spectrum and the
chemical composition of the cosmic ray flux above several hundred TeV, and these
capabilities allow for additional studies of the anisotropy.  For one of the
excess regions observed in the 10\,TeV skymap, the Milagro experiment has reported 
a different energy spectrum than the isotropic cosmic ray flux \citep{Abdo:2008kr}. 
With data from IceTop, studies of the energy spectrum and composition of the cosmic 
ray flux in distinct regions of the southern sky can be performed.

IceTop is now in stable running mode in its complete configuration of 81 stations.
In two years, the size of the cosmic ray data set available for anisotropy studies will 
be more than twice what was used in the analysis presented in this paper.  Eventually, 
it will be possible to extend the analysis of cosmic ray anisotropy to higher energies.

%************************************************************************************
\section{Acknowledgments}\label{sec:Acknowledgments}
%************************************************************************************

\acknowledgments

Some of the results in this paper have been derived using the
HEALPix~\citep{Gorski:2004by} software libraries.

We acknowledge the support from the following agencies:
U.S. National Science Foundation-Office of Polar Programs,
U.S. National Science Foundation-Physics Division,
University of Wisconsin Alumni Research Foundation,
the Grid Laboratory Of Wisconsin (GLOW) grid infrastructure at the University of Wisconsin - Madison, the Open Science Grid (OSG) grid infrastructure;
U.S. Department of Energy, and National Energy Research Scientific Computing Center,
the Louisiana Optical Network Initiative (LONI) grid computing resources;
National Science and Engineering Research Council of Canada;
Swedish Research Council,
Swedish Polar Research Secretariat,
Swedish National Infrastructure for Computing (SNIC),
and Knut and Alice Wallenberg Foundation, Sweden;
German Ministry for Education and Research (BMBF),
Deutsche Forschungsgemeinschaft (DFG),
Research Department of Plasmas with Complex Interactions (Bochum), Germany;
Fund for Scientific Research (FNRS-FWO),
FWO Odysseus programme,
Flanders Institute to encourage scientific and technological research in industry (IWT),
Belgian Federal Science Policy Office (Belspo);
University of Oxford, United Kingdom;
Marsden Fund, New Zealand;
Australian Research Council;
Japan Society for Promotion of Science (JSPS);
the Swiss National Science Foundation (SNSF), Switzerland.

%************************************************************************************

\end{document}